# Measuring hand use in the home after cervical spinal cord injury using egocentric video

Andrea Bandini[1,2,3], Mehdy Dousty[1,4], Sander L. Hitzig[5,6,7], B. Catharine Craven[1,8,9], Sukhvinder Kalsi-Ryan[1,10], José Zariffa[1,4,5,11] *

1. KITE Research Institute, Toronto Rehabilitation Institute – University Health Network, Canada
2. The BioRobotics Institute, Scuola Superiore Sant'Anna, Pisa, Italy
3. Department of Excellence in Robotics and AI, Scuola Superiore Sant'Anna, Pisa, Italy
4. Institute of Biomedical Engineering, University of Toronto, Toronto, ON, Canada
5. Rehabilitation Sciences Institute, Temerty Faculty of Medicine, University of Toronto, Toronto, ON, Canada
6. St. John's Rehab Research Program, Sunnybrook Research Institute, Sunnybrook Health Sciences Centre, Toronto, ON, Canada
7. Department of Occupational Science & Occupational Therapy, Temerty Faculty of Medicine, University of Toronto, Toronto, ON, Canada
8. Brain and Spinal Cord Rehabilitation Program, Toronto Rehabilitation Institute – University Health Network, Toronto, Ontario, Canada
9. Division of Physical Medicine and Rehabilitation, Department of Medicine, Temerty Faculty of Medicine, University of Toronto, Toronto, ON, Canada
10. Department of Physical Therapy, University of Toronto, Toronto, ON, Canada
11. Edward S. Rogers Sr. Department of Electrical and Computer Engineering, University of Toronto, Toronto, ON, Canada

**\* Corresponding author**
Jose Zariffa, PhD, Peng
Email: jose.zariffa@utoronto.ca
Tel: +1 416-597-3422 x7915
Office: The KITE Research Institute, University Health Network, 550 University Avenue, #12-102, Toronto, ON, M5G 2A2 Canada

**Abstract**

Egocentric video has recently emerged as a potential solution for monitoring hand function in individuals living with tetraplegia in the community, especially for its ability to detect functional use in the home environment. The aim of this study was to develop and validate a wearable vision-based system for measuring hand use in the home among individuals living with tetraplegia.

Several deep learning algorithms for detecting functional hand-object interactions were developed and compared. The most accurate algorithm was used to extract measures of hand function from 65 hours of unscripted video recorded at home by 20 participants with tetraplegia. These measures were: the percentage of interaction time over total recording time (*Perc*); the average duration of individual interactions (*Dur*); the number of interactions per hour (*Num*). To demonstrate the clinical validity of the technology, egocentric measures were correlated with validated clinical assessments of hand function and independence (Graded Redefined Assessment of Strength, Sensibility and Prehension - *GRASSP*, Upper Extremity Motor Score - *UEMS*, and Spinal Cord Independent Measure - *SCIM*).

Hand-object interactions were automatically detected with a median F1-score of 0.80 (0.67-0.87). Our results demonstrated that higher *UEMS* and better prehension were related to greater time spent interacting, whereas higher *SCIM* and better hand sensation resulted in a higher number of interactions performed during the egocentric video recordings. For the first time, measures of hand function automatically estimated in an unconstrained environment in individuals with tetraplegia have been validated against internationally accepted measures of hand function. Future work will necessitate a formal evaluation of the reliability and responsiveness of the egocentric-based performance measures for hand use.

## 1. INTRODUCTION

The functional use of the upper extremities (UE) is the top recovery priority for most individuals with cervical spinal cord injury (cSCI).[1,2] However, optimizing the recovery of UE function is complicated by the fact that improvements observed in the clinic do not translate into increased hand use in the community.[3–5] Furthermore, assessment beyond the clinical setting is crucial because of current financial pressures in the healthcare system, which results in patients often being discharged although they are at a phase in their recovery where they may still experience large improvements.[6–8] After discharge from the inpatient rehabilitation setting, the distance between one's home and the clinical settings can complicate the accurate tracking of neurologic and functional recovery.[9] These barriers to optimal UE rehabilitation have a negative impact on an individual's ability to regain independence in their daily life at home.

To address this gap, rehabilitative interventions should be targeted towards improving UE use during daily life in the community.[5] Therefore, it is imperative to develop and validate outcome measures of UE function that capture the *performance* domain of the International Classification of Functioning, Disability and Health,[10] which refers to how individuals with cSCI carry out activities of daily living (ADLs) and instrumental ADLs (iADLs) in their current environment.[11] The ideal solution should: (1) be able to track UE recovery remotely; (2) produce useful outcome measures for planning and validating interventions; and (3) be able to measure UE activity performance[12] in the individual's free-living environment[4,5] in order to understand whether improvements in UE motor function translate into an increased use in daily life.

Over the past 10 years, several approaches based on wearable sensors have been proposed.[3,13–20] Accelerometers and inertial measurement units have largely been used for monitoring UE functions, with their small size and capacity to record UE kinematics for several hours making them strong candidates for measuring UE function in an ecologically valid manner.[3,13–15,18] However, these devices cannot provide information about finger movements, grasp types, and the context of functional hand use (i.e., what type of object is the hand in contact with? what type of action is being conducted?). Finger-worn accelerometers[19,20] and sensorized gloves[17] may provide more details about hand and finger movements, but it remains difficult to understand the context of functional use of the hand. Moreover, they may be cumbersome for people with impaired function and sensation. Other approaches based on magnetic sensors[16] may instead be sensitive to metal objects commonly used in the household (e.g., kitchen utensils, doorknobs, etc.) that could limit their use in unconstrained environments.

Egocentric vision (i.e., approaches based on wearable cameras) has recently emerged as an alternative solution to monitoring UE functions in individuals with cSCI and stroke survivors.[21–26] When the camera is worn on the head, camera movements are driven by the user's attention, with the intrinsic advantage of focusing on the hands and manipulated objects.[23,27] Moreover, this is the only technology that provides functional context of hand use. From the egocentric perspective, it is possible not only to automatically detect hands and objects,[23,28,29] but also what types of actions and activities are being conducted.[30–32]

For the purpose of monitoring UE function in individuals with cSCI, we recently developed computer vision algorithms for detecting functional hand-object interactions,[21,33,34] as the presence of such interactions can constitute the basis for novel outcome measures reflecting the performance domain of hand function. Egocentric measures of hand function recently proposed by Likitlersuang *et al.* (2019)[21] are the percentage of interaction over total recording time, the average duration of interactions, and the number of interactions per hour. However, these measures have never been extracted from the individuals' free-living environment (e.g., at home during unscripted activities), nor have they been correlated with clinic-based assessments of UE function and SCI independence. These are essential requirements for demonstrating the clinical validity of the technology and towards translating it into outpatient care.

Thus, the overall goal of this work was to develop and validate a wearable vision-based system for measuring hand use in the home in individuals with cSCI. The main contributions of this manuscript are: (1) extensive algorithmic comparison to identify the most accurate approach for detecting hand-object interactions; (2) the computation of egocentric measures of hand function from more than 65 hours of unscripted video and their validation with clinic-based assessments.

## 2. METHODS
### 2.1 Participants and data collection
The study was approved by the Research Ethics Boards at the UHN – Toronto Rehabilitation Institute. All participants and any household members appearing in the videos signed informed consent according to the requirements of the Declaration of Helsinki. The inclusion criteria were: age greater than 18 years; neurological level of injury between C3 and C8; AIS grade A-D; impaired but not completely absent hand function; access to a caregiver that can help with donning and doffing of the camera system (if the participant required help); to be able to turn off the wearable camera on their own, using the tablet provided. Exclusion criteria were: presence of other neuromusculoskeletal disease affecting upper limb movements; deformity of the upper limb joints; pain when moving the upper limbs.

The number of participants to be recruited was determined via power analysis. For an estimated correlation coefficient of 0.6 between the egocentric measures and clinical scores, a desired power of 0.8, and α-level of 0.05, the number of participants needed for the study was found to be 15.[35] We decided to use a sample size of 20 to include a minimum number of women (N=5) and to account for dropouts. Although sex and gender were not considered in the statistical analysis due to sample size constraints, we wanted to ensure that different gender perspectives were reflected in the type of activities conducted in the home.

Participants were invited to the rehabilitation center for a first meeting where they were assessed using the Spinal Cord Independence Measure III (*SCIM*)[36] and the Graded Redefined Assessment of Strength, Sensibility and Prehension (*GRASSP*).[37] The upper extremity motor score (*UEMS*) for both UEs was extrapolated from the *GRASSP*. The International Standards for Neurological Classification of SCI (*ISNCSCI*)[38] were extracted from clinical charts when available or self-reported.

A detailed explanation of the recording protocol is described in Tsai *et al.*, 2020.[39] Briefly, participants agreed to record their normal daily routine at home using a head-mounted camera (GoPro® Hero5 Black). They were asked to record three sessions of approximately 1.5 hours each, over a two-week period, that captured them performing ADLs or iADLs. To ensure the recording of naturalistic behaviors, we negotiated a set schedule with the participants on when recording activities of their normal routine would occur, while avoiding privacy issues. The recording schedule was usually composed of three time windows (e.g., Monday: 8-9:30 AM, Wednesday: 12-1:30 PM, Friday: 6-7:30 PM) selected in collaboration with each participant after a short interview to gather more information about their daily routine. The time schedule was subject-specific and was devised to meet the following requirements: (1) Presence of diversified activities within each time frame; (2) Absence of other people or reduced risk to record other people; (3) Absence of activities that may cause privacy issues; and (4) Recording duration of at least one consecutive hour to capture more natural behaviors.[26,39] Suggesting time frames instead of specific activities was previously found beneficial in preventing the recording of artificial repetitions of specific hand tasks.[40]

### 2.2 Video processing

We quantified hand use with an algorithmic framework that identifies the participant's hands[23,41] (i.e., hand localization) and predicts whether the detected hands interact with objects in a functional manner (i.e., interaction detection).[21,34] The outputs of this process are two binary temporal profiles (one for each hand) that, for each video frame, indicate whether the participant's hands are functionally interacting (score=1) or not (score=0).[21,23] From this profile we extracted the egocentric measures of hand function[21] as illustrated in Section 2.3. An overview of the video processing pipeline is illustrated in Fig. 1.

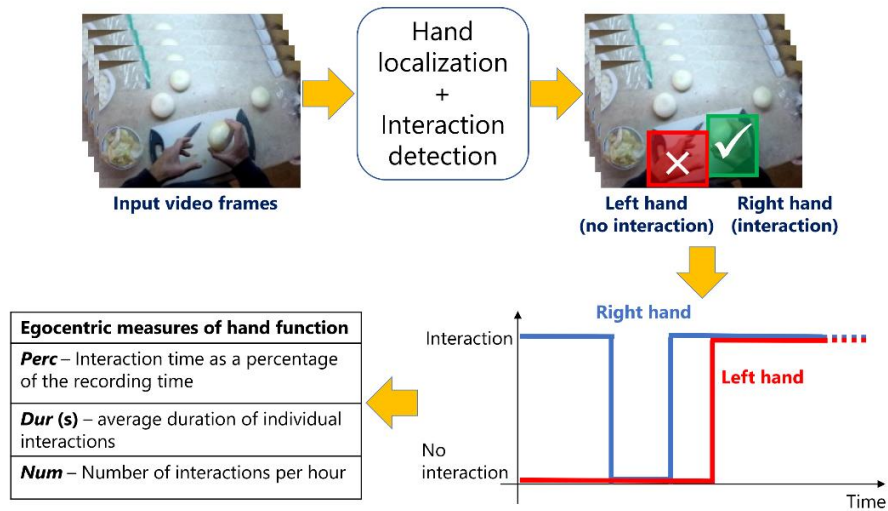

**Figure 1. Video processing pipeline developed for detecting hand-object interactions and extracting egocentric measures of hand function.**

**Hand localization** was performed with the deep-learning model recently proposed by Shan *et al.*, 2020.[42] For each frame, this model identifies the visible hands along with their side, the contact state of each hand (0=no contact, 1=self-contact, 2=contact with other person, 3=contact with non-portable object, 4=contact with portable object), as well as the location of the object the hand is in contact with. This model uses a Faster-RCNN architecture that localizes two objects (hands and contacted objects) with two additional fully connected layers for estimating the hand side and contact state.[42,43]

***Interaction detection***. On top of the hand localization, we predicted whether each detected hand was functionally interacting with an object. For the interaction detection, we implemented and compared multiple algorithms, eventually choosing the most accurate one to extract and validate the egocentric measures of hand function (Sec. 2.3).

To conduct a comparison among all algorithms, a trained annotator labelled the presence of functional hand-object interactions for both hands on a subset of frames. Specifically, we compared the following approaches:

1. *Contact state - frame-by-frame (**State$_{frame}$**)* - Shan's model already provides a contact state that indicates what type of contact exists between the detected hand and an object. We hypothesized that the contact state would constitute a proxy for functional interactions. Thus, for each frame, contact states $\geq 3$ (i.e., contact with portable and non-portable objects) were considered as functional interaction, whereas the rest of the states were considered as absence of functional interaction.

2. *Hand state with temporal pooling (**State$_{pool}$**)* - Like the above strategy, we considered contact states $\geq 3$ as functional interactions. Every 30 consecutive frames (i.e., 1 second) we estimated the interaction score using the majority vote among all frames within this time interval.

3. *Artificial neural network with Shan's model features (**Shan+ANN**)* - For each frame and each detected hand, we used the nine features obtained with Shan's model (four bounding box coordinates, one confidence score of hand localization, one hand-object contact state, and three offset values) as input to a custom artificial neural network (ANN) with two fully connected layers (32 and 16 units, respectively) followed by two output nodes for the predictions.

4. *Shan+ANN with temporal pooling (**Shan+ANN$_{pool}$**)* - The output of the above approach was pooled every 30 consecutive frames, by using the majority vote among the frames within the temporal window.

5. *Recurrent neural networks (RNNs) with Shan's model features* - The nine output features obtained for each hand with Shan's model were used as input to different RNN models. For all the models reported below, we used a time-window of 30 consecutive frames. Specifically, we used:

   a. A two-layer gated recurrent unit model (**Shan+GRU**) where each GRU layer had 9 hidden units. On top of the GRU layers, we stacked two fully connected layers with 16 and 8 units, respectively, followed by 2 output nodes for the predictions.

   b. A two-layer Long-Short-Term-Memory model (**Shan+LSTM**) - each LSTM layer had 9 hidden units. On top of the LSTM layers, we stacked two fully connected

  layers with 16 and 8 units, respectively, followed by 2 output nodes for the predictions.

 c. A two-layer bidirectional LSTM model (**Shan+BiLSTM**) - each LSTM layer had 9 hidden units. On top of the BiLSTM layers, we stacked two fully connected layers with 16 and 8 units, respectively, followed by 2 output nodes for the predictions.

The above approaches (except for 1 and 2, for which we used the original pre-trained model[42]) were trained and tested in a leave-one-subject-out cross-validation (LOSO-CV) fashion. Additional details about the strategies adopted for training the algorithms are reported in the supplemental material. Classification performance was calculated using the F1-score and compared with two baseline approaches. The first one, proposed by Likitlersuang *et al.* (2019),[21] used color, edge, and optical flow features extracted from the hand, background, and manipulated object to estimate the presence of hand-object interactions via a random forest classifier (**Likitlersuang model**). The second one used MobileNet V1 as a CNN architecture to classify the detected hand regions as interacting or non-interacting (Hand Object Interaction Detection Network - **HOID-Net**).[34]

### 2.3 Egocentric measures of hand function

The best model from section 2.2 (i.e., the one with highest F1-score) was implemented on the entire dataset to estimate egocentric measures of hand function and correlate them with the clinical scores obtained at the beginning of the study (Sec. 2.1). The egocentric measures of hand function, previously introduced by Likitlersuang *et al.* (2019),[21] were:

- *Perc*: The amount of total interaction as a percentage of the recording time.
- *Dur*: The average duration of individual interactions in seconds.
- *Num*: The number of interactions per hour.

Each measure was calculated for the dominant ($Perc_{DH}$, $Dur_{DH}$, and $Num_{DH}$) and non-dominant hand ($Perc_{NH}$, $Dur_{NH}$, $Num_{NH}$), considering the hand dominance after the injury. The bilateral version of *Perc* ($Perc_{Bi}$) was calculated as the average between the dominant and non-dominant hand values, whereas bilateral versions for *Dur* and *Num* ($Dur_{Bi}$ and $Num_{Bi}$) were calculated as the sum of values from dominant and non-dominant hands.

### 2.4 Statistical analysis

A non-parametric Friedman test was conducted to test for differences between the F1-scores obtained with the interaction-detection models. In case of significant difference among the algorithms, multiple comparison test using the Dunn & Sidák's approach was conducted.[44] As an additional metric to compare the algorithm performance, we computed the percentage of participants for which the F1-score was higher than 0.8.

  Egocentric measures from the dominant and non-dominant hands were compared using the Wilcoxon signed rank test. Correlations between the egocentric measures and the clinical scores were estimated using the Spearman's correlation coefficient (Rho).

Bilateral egocentric measures were correlated with the bilateral clinical scores: *UEMS*, *SCIM* sub-scores (self-care - $SCIM_s$; respiration and sphincter management - $SCIM_{RS}$; mobility - $SCIM_M$) and total score ($SCIM_{TOT}$). Unilateral egocentric measures were correlated with the bilateral clinical scores and with the unilateral scores from the corresponding hand, namely: *UEMS*, *GRASSP* sub-scores (strength - *GR-Str*; dorsal sensation - *GR-Sens (dorsal)*; palmar sensation - *GR-Sens (palmar)*; total sensation - *GR-Sens*; prehension ability - *GR-PA*; prehension performance - *GR-PP*) and total score (*GR-tot*).

### *3. RESULTS*

Twenty-one adults with cSCI (16 M, 5 F; Neurological level of injury between C3 and C8; AIS grade A-D; Months since injury = 12-432) were recruited for the study. The average bilateral *UEMS* was 37.1 ± 11.1 with a unilateral *UEMS* significantly higher for the dominant hand ($UEMS_{dominant}$ 19.8 ± 5.5; $UEMS_{nondominant}$ 17.3 ± 6.2; Z = -2.61, *p* = 0.009). An overview of the demographic and clinical characteristics of the participants is reported in Table 1. Participant 4 was excluded from the remainder of the analysis, as he was not able to record any videos due to health issues.

In total, more than 65 hours of video were recorded (approximately 3,880 minutes), at a resolution of 1080p and 30 fps. The average video duration per participant was 204 ± 42 minutes, and an average of 65 ± 42 minutes of video were recorded per day. The total number of frames was approximately 7.35 million. For subsequent analyses, all frames were resized to 720 × 405 pixels. Examples of egocentric images captured during the experiments are shown in Fig. 2.

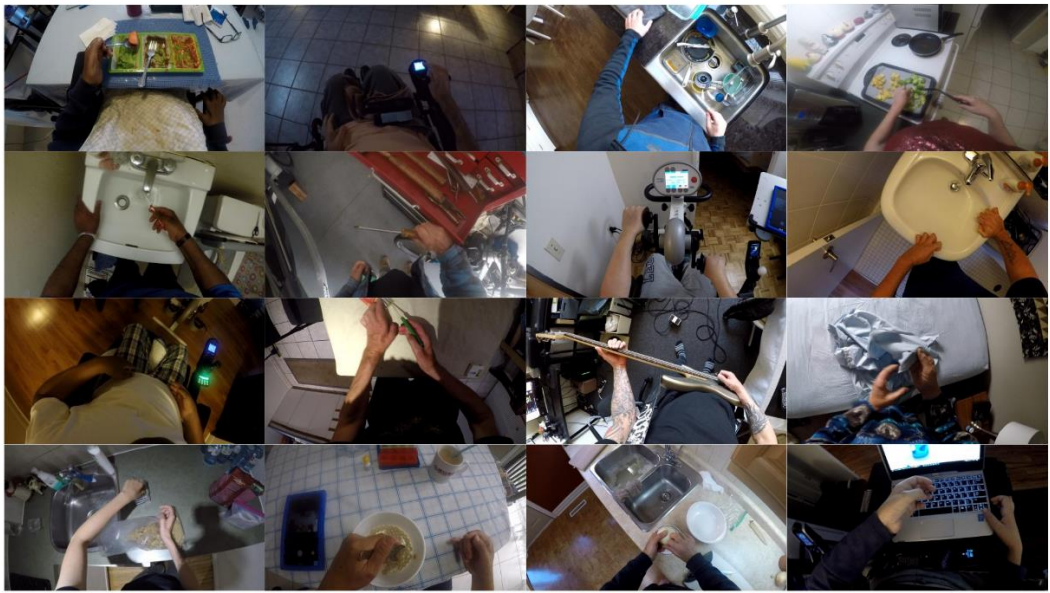

***Figure 2. Examples of egocentric frames collected by the participants of this study.***

**Table 1.** Demographic information for the participants recruited in this study. Injury type: T = Traumatic; NT = non-traumatic. Hand dominance: R = right; L = left. For the level of injury and AIS, self-reported values are indicated with an asterisk. § Participant n. 4 was recruited but did not complete the study.

| ID | Sex | Age (years) | Months from injury | Injury type | Level of injury | AIS | Dominant hand (pre-injury) | UEMS$_{dominant}$ | UEMS$_{nondominant}$ | UEMS$_{tot}$ |
|---|---|---|---|---|---|---|---|---|---|---|
| 1 | M | 46 | 57 | T | C5 | B* | L(R) | 7 | 7 | 14 |
| 2 | M | 61 | 60 | T | C4* | D* | L(L) | 23 | 25 | 48 |
| 3 | M | 54 | 53 | T | C4* | D* | L(R) | 24 | 18 | 42 |
| 4§ | M | 53 | 12 | NT | C6* | D* | R(R) | 21 | 24 | 45 |
| 5 | M | 44 | 25 | T | C5* | D* | R(R) | 25 | 23 | 48 |
| 6 | M | 63 | 51 | NT | C4* | D* | R(R) | 20 | 23 | 43 |
| 7 | M | 47 | 88 | T | C8 | D* | L(L) | 24 | 22 | 46 |
| 8 | M | 63 | 18 | T | C5 | D* | R(R) | 15 | 13 | 28 |
| 9 | M | 63 | 91 | T | C4 | A | R(L) | 18 | 12 | 30 |
| 10 | M | 61 | 13 | T | C3* | C* | R(L) | 25 | 14 | 39 |
| 11 | M | 49 | 240 | T | C5 | C | L(L) | 10 | 7 | 17 |
| 12 | M | 63 | 13 | T | C5 | D | L(L) | 25 | 24 | 49 |
| 13 | M | 54 | 31 | T | C3 | C | R(R) | 19 | 9 | 28 |
| 14 | F | 62 | 38 | T | C4 | D | R(R) | 19 | 18 | 37 |
| 15 | F | 21 | 17 | NT | C5 | D | R(R) | 23 | 25 | 48 |
| 16 | F | 32 | 63 | T | C7 | D | R(L) | 23 | 19 | 42 |
| 17 | M | 59 | 15 | T | C4* | D* | R(R) | 24 | 20 | 44 |
| 18 | M | 63 | 432 | T | C5* | B | R(L) | 10 | 9 | 19 |
| 19 | M | 26 | 31 | T | C6* | A | R(R) | 14 | 12 | 26 |
| 20 | F | 75 | 19 | T | C5 | D | R(L) | 22 | 17 | 39 |
| 21 | F | 67 | 49 | NT | C4 | D | R(R) | 24 | 23 | 47 |
| Mean ± SD | | 53.6 ± 13.8 | 67.4 ± 97.2 | | | | | 19.8 ± 5.5 | 17.3 ± 6.2 | 37.1 ± 11.1 |

### 3.1 Interaction detection performance

The interaction detection algorithms presented in section 2.2 were evaluated on 632,180 manually annotated frames obtained from 13 participants (IDs 1-3 and 5-14). A summary of the interaction detection performance is reported in Table 2. Friedman's test showed a significant difference among the eight algorithms. The multiple comparison test indicated significant differences between $State_{pool}$ and $Likitlersuang\ model$ ($p$=0.0021), $Shan+ANN_{pool}$ and $Likitlersuang\ model$ ($p$=0.0029), and between $Shan+LSTM$ and $Likitlersuang\ model$ ($p$=0.030). Thus, we can state that $State_{pool}$, $Shan+ANN_{pool}$, and $Shan+LSTM$ produced higher performance than $Likitlersuang\ model$. However, no significant differences were found among $State_{frame}$, $State_{pool}$, $Shan+ANN$, $Shan+ANN_{pool}$, $Shan+GRU$, $Shan+LSTM$, $Shan+BiLSTM$, and $HOID\text{-}Net$. As a tie-breaker rule, we used the percentage of participants for which the F1-score was higher than 0.8. With this criterion, $State_{pool}$ was the approach implemented on the entire dataset (~7.35 million frames) to extract the egocentric measures of hand function (Table 2). The processing time required to run $State_{pool}$ on the entire dataset was approximately 100 ms per frame, with non-optimized code, using an NVIDIA® GeForce RTX™ 2080 Ti 11GB GPU.

**Table 2.** Interaction detection performance of the algorithms tested in this work. The best algorithm is highlighted in bold

| Algorithm | F1-score Median (inter-quartile range) | Percentage of participants with F1-score > 0.8 |
|---|---|---|
| $State_{frame}$ | 0.80 (0.67-0.85) | 46.2 |
| **$State_{pool}$** | **0.80 (0.67-0.87)** | **53.8** |
| Shan+ANN | 0.78 (0.65-0.85) | 38.5 |
| $Shan+ANN_{pool}$ | 0.79 (0.66-0.86) | 38.5 |
| Shan+GRU | 0.74 (0.64-0.85) | 38.5 |
| Shan+LSTM | 0.77 (0.65-0.84) | 38.5 |
| Shan+BiLSTM | 0.75 (0.60-0.83) | 38.5 |
| HOID-Net | 0.71 (0.56-0.80) | 30.8 |
| Likitlersuang model | 0.77 (0.57-0.79) | 23.1 |
| Friedman's test: $X^2(8) = 24.77$, $p < .001$ | | |

### 3.2 Egocentric measures of hand function

**Dominant vs non-dominant hand**. According to Table 3, participants used their dominant hand significantly longer than the non-dominant one (i.e., larger values of $Perc$ and $Dur$ from the dominant hand). The average number of interactions, however, did not differ between the two hands.

**Table 3.** Comparison between the proposed outcome measures extracted from dominant and non-dominant hand. Outcome measures were reported with their median values and inter-quartile ranges.

|  | Dominant hand | Non-dominant hand | Wilcoxon Signed-Rank test |
|---|---|---|---|
| Perc | 0.65 (0.59-0.74) | 0.56 (0.41-0.68) | $Z = 2.65$; $p = 0.008$ |
| Dur (s) | 23.5 (15.5-32.0) | 18.1 (11.2-23.2) | $Z = 2.09$; $p = 0.037$ |
| Num | 102.0 (83.4-134.3) | 112.6 (90.5-130.2) | $Z = -0.63$; $p = 0.526$ |

***Egocentric measures vs bilateral clinical scores (Fig. 3a)***. In general, unilateral and bilateral values of *Perc* and *Num* showed moderate-to-strong positive correlations with the bilateral clinical scores. For the bilateral egocentric measures (Fig. 3a - left): $Perc_{Bi}$ showed moderate positive correlations with bilateral *UEMS*, *SCIM* and its subscores; $Num_{Bi}$ showed a strong positive correlation with $SCIM_M$, and moderate positive correlations with bilateral *UEMS*, $SCIM_{TOT}$, $SCIM_S$, and $SCIM_{RS}$; all correlations between $Dur_{Bi}$ and the bilateral clinical scores were very weak or negligible.

A similar pattern was also visible for the unilateral egocentric measures. Specifically, for the dominant hand (Fig 3a - center), $Perc_{DH}$ and $Num_{DH}$ showed weak-to-moderate positive correlations with bilateral *UEMS* and *SCIM* subscores, whereas all correlations with $Dur_{DH}$ were very weak or negligible. For the non-dominant hand (Fig 3a - right): $Perc_{NH}$ showed moderate positive correlations with bilateral *UEMS*, *SCIM* and its subscores; $Num_{NH}$ showed strong positive correlations with $SCIM_{TOT}$, $SCIM_M$, and $SCIM_{RS}$, and moderate positive correlations with bilateral *UEMS* and $SCIM_S$; all correlations with $Dur_{NH}$ were weak or very weak.

***Egocentric measures vs unilateral clinical scores (Fig. 3b)***. In general, *Perc* and *Num* showed positive correlations with *UEMS* and *GRASSP* items, with weak-to-moderate values for the dominant hand and moderate-to-strong values for the non-dominant hand.

For the dominant hand (Fig. 3b - left): $Perc_{DH}$ showed a moderate positive correlation with *GR-PP*, and weak positive correlations with unilateral *UEMS* and the other *GRASSP* items (except for *GR-Sens (dorsal)*); $Num_{DH}$ showed moderate positive correlations with *GRASSP* sensation items, and weak positive correlations with unilateral *UEMS* and the other *GRASSP* items; $Dur_{DH}$ showed weak negative correlations with the *GRASSP* sensation items, and very-weak or negligible correlations with the other scores.

For the non-dominant hand (Fig. 3b - right): $Perc_{NH}$ showed a strong positive correlation with *GR-PA*, and moderate positive correlations with *UEMS*, *GR-Str*, *GR-PP*, and *GR-tot* (moderate positive), whereas correlations with the *GRASSP* sensation scores were weak positive; $Num_{NH}$ showed moderate positive correlations with *UEMS* and most of the *GRASSP* susbscores; $Dur_{NH}$ showed a moderate positive correlation with *GR-PA*, weak positive correlations with *UEMS*, *GR-Str*, *GR-PP*, and *GR-tot*, and negligible correlations with the *GRASSP* sensation scores.

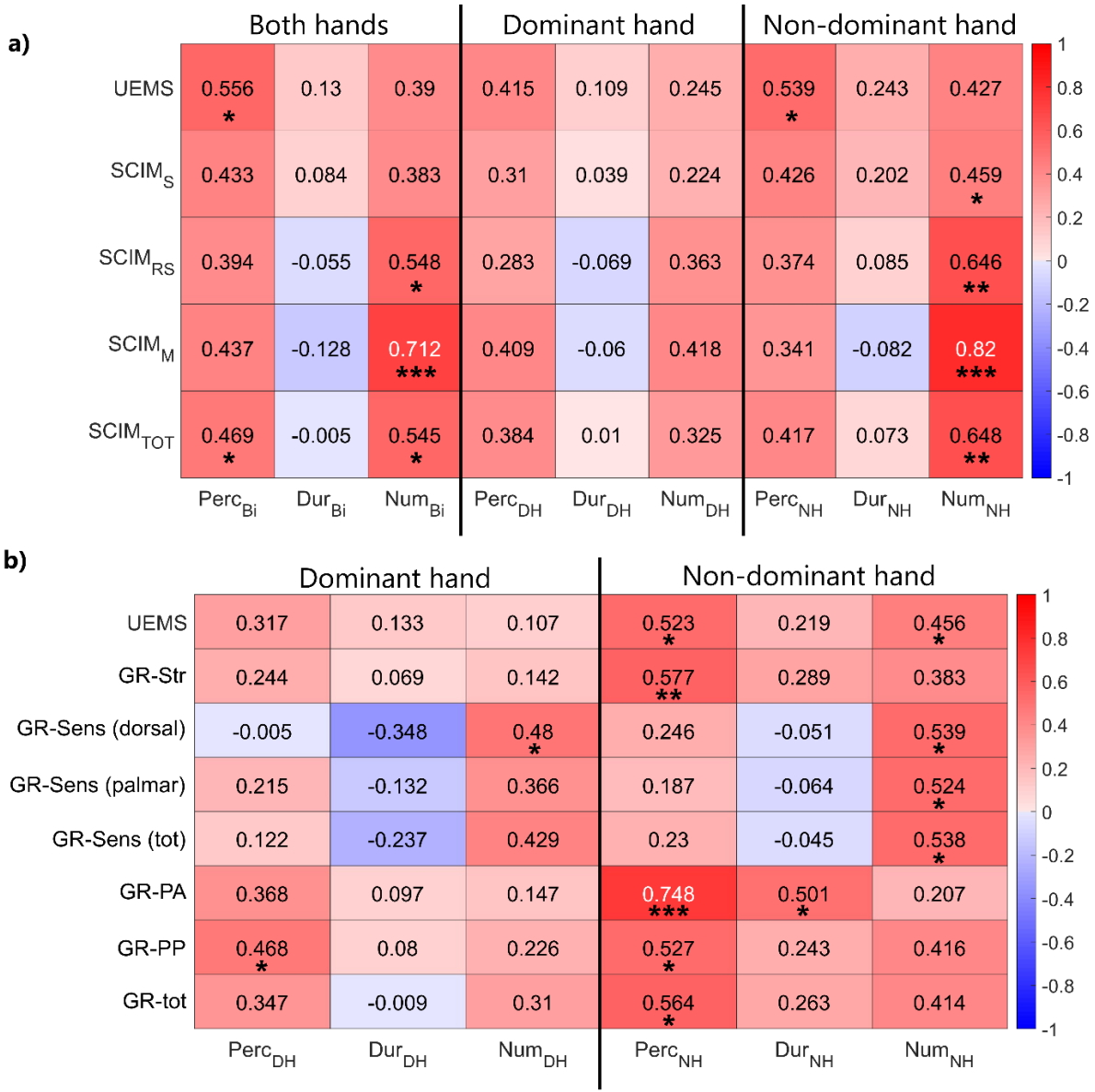

***Figure 3. Correlation heatmaps between the egocentric measures of hand function and the clinical scores collected before the recordings: a) correlations between egocentric measures and bilateral clinical scores (bilateral UEMS, SCIM, and SCIM subscores); b) correlations between egocentric measures and unilateral clinical scores (unilateral UEMS, GRASSP, and GRASSP subscores). \* p < .05, \*\* p < .01, \*\*\* p < .001.***

## 4. DISCUSSION

We proposed and validated a video-based algorithmic framework for measuring hand use in individuals with tetraplegia who recorded activities of their daily routine in the community using a wearable camera. To the best of our knowledge, this is the first time that measures of hand function automatically estimated in an unconstrained environment after cSCI have been validated against clinic-based assessments.

### 4.1 Hand-object interaction detection

In the proposed approach, the automated detection of hand-object functional interactions was the basis for extracting egocentric measures of hand function. We conducted extensive algorithm comparison to detect the most accurate approach among those described in section 2.2. Results highlighted similar performance among the approaches (see Tab. 2) suggesting that the use of the hand-object contact state can be used as a proxy for estimating functional interactions. Temporal pooling introduced a slight yet meaningful performance boost. This improvement can be due to the smoothing effect that temporal pooling has, which removes short and isolated sequences of frames that may contain wrong predictions. This result is in line with the approach proposed by Likitlersuang *et al.* (2019),[21] where the raw interaction detection results were processed with a moving average filter, and further corroborates the hypothesis that temporal information must be included in the analysis of functional hand-object interactions.

### 4.2 Egocentric measures of hand function

As illustrated in Table 3, we observed that participants used their dominant hand more than the non-dominant one, as highlighted by the higher percentage of time spent interacting. However, since the number of interactions was not significantly different between the two hands, we can state that the interactions performed with the dominant hand were - on average - longer than those conducted with the non-dominant one.

Correlations between egocentric measures and bilateral clinical scores (Fig. 3a) suggested that individuals with higher *UEMS* spent more time conducting functional interactions, whereas higher *SCIM* corresponded to higher percentage of time spent interacting and higher number of interactions. These results suggest that participants with a higher level of independence not only spent more time using their hands but also performed more interactions. Considering the presence of a strong positive correlation between $Num_{Bi}$ and $SCIM_M$, we believe that increased mobility could have helped them conduct a higher number of activities during the experiments. Higher *UEMS*, however, may have some relationship only with the percentage of time spent using the hands. This result can be explained by the fact that participants with lower *UEMS*, many of whom who have increased muscle weakness, may be exposed to muscle fatigue during activities that could prevent them from engaging in longer periods of interactions, as compared to participants with higher *UEMS*.

Like the bilateral clinical scores, correlations with unilateral scores were more visible in the non-dominant hand. Specifically, participants with higher *UEMS* in the non-dominant hand spent more time interacting and conducted more interactions with this hand. This result was less visible in the dominant hand. Similarly, higher *GR-tot* corresponded to more time spent interacting, with higher correlations in the non-dominant hand. By analyzing the correlations with *GRASSP* sub-scores, we observed the following behaviors: (1) Higher strength resulted in more time spent interacting (especially in the non-dominant hand) - considering that *UEMS* is a sub-score of *GR-Str*, this result may further suggest the presence of fatigue in individuals with lower strength; (2) Higher sensation resulted in a higher number of interactions (both hands) - also, there were weak negative correlations between $Dur_{DH}$ and the sensation sub-scores from the dominant hand, which may suggest that better sensation may also shorten the duration of functional interactions conducted with the dominant hand; (3) Better prehension led to more time spent interacting (both hands) and longer interactions (only for the non-dominant hand). These findings suggest that sensation and prehension can be two key-factors related to hand usage, with the former related to an increase of the time spent interacting, and the latter associated with a higher number of interactions.

The inconsistencies between the dominant and non-dominant hands in terms  may be explained by the fact that 7 out of 20 participants changed hand dominance because of the injury (see Tab. 1). Although the reduced sample size makes it hard to state that one correlation was significantly larger or smaller than another, we believe that the higher correlations obtained in the non-dominant hand can be explained by the fact that the dominant hand was the most used one. This fact may have led participants to also adopt compensatory strategies to conduct activities despite the injury. Instead, the consequences of the injury could be more visible in the non-dominant hand, as this was the least used one. However, the fact that the dominant hand was used more than the non-dominant one does not necessarily mean that it was used correctly (i.e., using the right grasp type). In fact, the better the prehension, the higher the percentage of time spent interacting, as demonstrated by the moderate positive correlation between *GR-PP* and $Perc_{DH}$ (Fig. 3b). Therefore, in addition to measuring the quantity of hand usage, the analysis of quality of hand usage (e.g., with grasp recognition techniques and by identifying hand roles during bimanual manipulations [24,45]) will also be of paramount importance to provide a comprehensive picture of how individuals with cSCI use their hands at home. Egocentric video provides a promising platform for these investigations.

Although this is the first time that measures of hand function obtained from egocentric vision have been validated against internationally accepted measures of hand function in cSCI, the magnitude of the correlations are approximately what is seen in other neurological populations comparing in-clinic to out-of-clinic assessments. An example is the research conducted by Barth *et al* (2020),[46] where moderate correlations between accelerometry measures of UE function and compensatory movement score were observed in individuals post-stroke both in-clinic and out-of-clinic conditions. Hence, our work constitutes another step towards validating wearable technologies for monitoring UE function in people with neurological impairments.

The correlations with *UEMS* and *GRASSP* scores demonstrated that the percentage of interaction time and number of functional interactions per hour may provide important information on how UE motor functions translate into an increased UE usage in daily life. This is an important finding towards proposing novel outcome measures of hand function able to measure performance in addition to capacity according to the terminology of the International Classification of Functioning, Disability and Health.[12]

### 4.3 Limitations and future work

Although we used the hand-object contact state for inferring functional interactions, its meaning only approximates functional hand-object interactions and further reasoning will be required to accurately predict the functional use of the hands from video-data. As an example, we will explore the use of action and activity recognition algorithms[30,31] with two goals in mind: (1) to develop interaction detection algorithms robust to the type of activities conducted at home and (2) to understand which actions require more intervention during rehabilitation. The limited sample size and the reduced number of female participants did not allow us to carry out any sex-based analysis. Moreover, there was a high inter-subject variability in our dataset that could affect the performance of the interaction detection step.

In addition to expanding the dataset, in future studies, additional aspects of UE impairment should be captured, including spasticity and tone. Egocentric vision may allow for collection of spasm frequency and identify when a person's UE is locked in a flexor or extensor synergy. However, additional work with computer vision and machine learning algorithms to identify these events and distinguish them clinically is needed. We believe that the possibility to capture a comprehensive picture of the UE status will help develop personalized algorithms, improve the classification performance, and foster the translation of this approach into clinical practice.

Another limitation of this study was the fact that the ISNCSCI was not collected at the time of the experiments. Thus, there could be a difference between the actual ISNCSCI at the time of the experiments and the one obtained in the charts or self-reported. However, considering that all the clinical scores used for validating the egocentric measures of hand function (i.e., *GRASSP*, *SCIM*, and *UEMS*) were obtained during the first visit, we believe that this limitation did not influence the results.

Whatever the algorithmic improvements will be, the processing pipeline developed in this project is modular and consists of two main blocks (hand localization and interaction detection). This feature will give us flexibility to accommodate evolving algorithms without changing the conceptual skeleton of the video-processing. While the present work demonstrates the validity of the egocentric-based performance measures for hand use, further work will be required to establish their reliability and responsiveness.

## 5. CONCLUSIONS

We demonstrated that computer vision and wearable cameras can be used to quantify hand use in individuals with tetraplegia living in the community. The detection of hand-object interactions constitutes the basis for extracting novel outcome measures of hand function. Specifically, we established validity of *Perc* and *Num* as measures of hand function at home to capture the performance domain of UE use. Our results demonstrated that higher *UEMS* and better prehension were related to more time spent interacting (higher *Perc*), whereas higher *SCIM* and better hand sensation resulted in a higher number of interactions performed during the experiments. Future research will validate these findings also in stroke survivors living in the community. Moreover, considering that the proposed measures provide information only about quantity of hand use, it will be interesting to link the proposed approach with the automated detection of grasp types, in order to detect quality of hand usage and propose a multi-dimensional model of outcome measures of hand function.

## ACKNOWLEDGMENTS

This work was supported by the Craig H. Neilsen Foundation (grant number 542675). The authors would also like to thank all the participants and their families involved in the study.

## DATA AVAILABILITY STATEMENT

Portions of the dataset will be shared for academic purposes upon reasonable request (see the supplemental material for more information). Please contact the corresponding author for data requests.

## CONFLICTS OF INTEREST

SKR is the CEO of Neural Outcomes Consulting Inc. which is the manufacturer of GRASSP products. The other authors report no conflicts of interest.